\documentclass[apjl]{emulateapj}

\usepackage{graphicx}

\def\mt{}
\def\vv{ }
\def\es{  }
\def\esn{  }
\def\mtt{  }
\def\mttt{  }
\def\ess{ }

\def\be{\begin{equation}}
\def\en{\end{equation}}

 \slugcomment{Accepted for pubblication in \textit{ApJL}}

 \shortauthors{Striani et al. 2011}

\begin{document}

\title{The Crab Nebula super-flare in April 2011: extremely fast particle acceleration and gamma-ray emission}




\author{E.~Striani\altaffilmark{2}, M.~Tavani\altaffilmark{1,2},
  G.~Piano\altaffilmark{1},
  I.~Donnarumma\altaffilmark{1},
G.~Pucella\altaffilmark{13}, V.~Vittorini\altaffilmark{1},
A.~Bulgarelli\altaffilmark{9}, A.~Trois\altaffilmark{1},
C.~Pittori\altaffilmark{14}, F. Verrecchia\altaffilmark{14},
E.~Costa\altaffilmark{1}, M. Weisskopf\altaffilmark{16},
A.~Tennant\altaffilmark{16},
 A.~Argan\altaffilmark{1},
G.~Barbiellini\altaffilmark{6},
P.~Caraveo\altaffilmark{3}, M~Cardillo\altaffilmark{1,2}
 P.~W.~Cattaneo\altaffilmark{7},
A.~W.~Chen\altaffilmark{3,4},
 G.~De
Paris\altaffilmark{1}, E.~Del Monte\altaffilmark{1},
G.~Di~Cocco\altaffilmark{5},  Y.~Evangelista\altaffilmark{1},
A.~Ferrari\altaffilmark{4,17}
M.~Feroci\altaffilmark{1}, 
F.~Fuschino\altaffilmark{5}, M.~Galli\altaffilmark{8},
F.~Gianotti\altaffilmark{5}, A. Giuliani\altaffilmark{3},
C.~Labanti\altaffilmark{5}, I.~Lapshov\altaffilmark{1},
F.~Lazzarotto\altaffilmark{1}, F.~Longo\altaffilmark{6},
M.~Marisaldi\altaffilmark{5}, S. Mereghetti\altaffilmark{3},
A.~Morselli\altaffilmark{11}, L.~Pacciani\altaffilmark{1},
A.~Pellizzoni\altaffilmark{18}, F.~Perotti\altaffilmark{3},
P.~Picozza\altaffilmark{2,11}, M.~Pilia\altaffilmark{18},
 M.~Rapisarda\altaffilmark{13},
A.~Rappoldi\altaffilmark{7}, S.~Sabatini\altaffilmark{1},
P.~Soffitta\altaffilmark{1}, M.~Trifoglio\altaffilmark{5},
 S. Vercellone\altaffilmark{15},
F. Lucarelli\altaffilmark{14}, P.~Santolamazza\altaffilmark{14},
P.~Giommi\altaffilmark{14} }

\altaffiltext{1} {INAF/IASF-Roma, I-00133 Roma, Italy}
\altaffiltext{2} {Dip. di Fisica, Univ. Tor Vergata, I-00133 Roma,
Italy} \altaffiltext{3} {INAF/IASF-Milano, I-20133 Milano, Italy}
\altaffiltext{4} {CIFS-Torino, I-10133 Torino, Italy}
\altaffiltext{5} {INAF/IASF-Bologna, I-40129 Bologna, Italy}
\altaffiltext{6} {Dip. Fisica and INFN Trieste, I-34127 Trieste,
Italy} \altaffiltext{7} {INFN-Pavia, I-27100 Pavia, Italy}
\altaffiltext{8} {ENEA-Bologna, I-40129 Bologna, Italy}
\altaffiltext{9} {INFN-Roma La Sapienza, I-00185 Roma, Italy}
\altaffiltext{10} {CNR-IMIP, Roma, Italy} \altaffiltext{11} {INFN
Roma Tor Vergata, I-00133 Roma, Italy} \altaffiltext{12} {Dip. di
Fisica, Univ. Dell'Insubria, I-22100 Como, Italy}
\altaffiltext{13} {ENEA Frascati,  I-00044 Frascati (Roma), Italy}
\altaffiltext{14} {ASI Science Data Center, I-00044
Frascati(Roma), Italy}  \altaffiltext{15} {INAF-IASF
Palermo, Palermo, Italy}
\altaffiltext{16} {NASA, Marshall  Space Flight Center,
Huntsville, AL 36812 (USA)} \altaffiltext{17} {Dip. Fisica,
Universit\'a di Torino, Turin, Italy}
\altaffiltext{18} {INAF-Osservatorio Astronomico di
Cagliari, localita' Poggio dei Pini, strada 54, I-09012 Capoterra,
Italy}



\begin{abstract}

We report on
{\mtt the} extremely intense and fast gamma-ray flare above 100
MeV detected by AGILE from the Crab Nebula  in mid-April 2011.
This {\mtt event is the fourth}
of a sequence of {\mt reported major} gamma-ray flares produced by
the Crab Nebula in the period 2007/mid-2011.
These  events are attributed to strong radiative {\mtt and plasma}
instabilities in the inner Crab Nebula, and
their properties are crucial {\mtt for theoretical studies of fast
and efficient particle acceleration up to $10^{15} \, \rm eV$.}
Here we study the very rapid flux and spectral evolution of the
event that reached {\mtt on April 16, 2011} the record-high peak
flux of $F = ({\esn 26 \pm 5}) \times 10^{-6} \rm \, ph \, cm^{-2}
\, s^{-1}$ {\mtt with a risetime timescale that we determine to be
in the range 6-10 hrs.}
{\mtt The peak flaring gamma-ray spectrum
reaches a distinct maximum near 500 MeV with no
substantial emission above 1 GeV.} The very rapid {\mtt risetime
and overall evolution} of the Crab Nebula flare strongly constrain
the acceleration mechanisms {\mtt and challenge MHD models.}
 We briefly discuss the
 theoretical implications of our {\mtt observations}.

\end{abstract}



\maketitle


\section{Introduction}

The Crab Nebula  is a most remarkable system,
 consisting of a
rotationally-powered pulsar {\mtt of large spindown luminosity
($L_{sd} \simeq 5 \times 10^{38} \, \rm erg \, s^{-1} $)}
interacting with a surrounding nebula {\mtt at the center of the}
SN1054 supernova remnant
 {\mtt (see, e.g., Hester 2008, for a review of the Crab properties)}.
The inner nebula is energized by the powerful wave/particle output
from the pulsar, and shows distinctive optical and X-ray
brightness enhancements (``wisps'', ``knots'', and the ``anvil''
aligned with the pulsar ``jet'') (Scargle 1969; Hester 1995, 2002,
2008; Weisskopf 2000).
{\mtt Despite the small-scale optical and X-ray variations
detected on timescales of weeks-months, the overall high-energy
flux resulting from the unpulsed synchrotron radiation of the
inner Nebula has been considered essentially stable for many
decades.}

The discovery by the AGILE satellite of a strong gamma-ray flare
above 100 MeV from the Crab Nebula in September 2010
\cite{tavani1,tavani3} and the confirmation by the Fermi-LAT
\cite{buehler,abdo2} started a new era of investigation of the
Crab Nebula and of the particle acceleration processes in general.
 Three
 {\mtt intense}  gamma-ray
flaring episodes from the Crab Nebula have been reported\footnote{
{\esn Enhanced TeV emission from the Crab nebula has also been
reported by the air-shower ARGO-YBJ experiment in coincidence with
the September 2010 event (Aielli et al. 2010). However, this claim
 was not supported by the simultaneous observations by VERITAS
 \cite{ong}
and MAGIC \cite{mariotti}.}} {\mtt in the gamma-ray energy range
100 MeV - a few GeV  by AGILE and Fermi-LAT} {\esn prior to}
{\mtt April}  2011 (Tavani et al. 2011, hereafter T11; Abdo et al.
2011, hereafter A11).
%
 This activity has been attributed to transient emission in
the inner Nebula due to the lack of any variation in the pulsed
(radio and high-energy) signal of the Crab pulsar or of any
detectable alternative counterpart (e.g., Heinke 2010, Cusumano et
al. 2011) .



Starting on 2011 April 11-12, a new gamma-ray flaring episode with
substantial emission  above 100 MeV was {\ess first} detected by Fermi-LAT
\cite{atel-1,atel-3} and {\ess then confirmed by} AGILE \cite{atel-2}. The flare developed
in the following days with substantial gamma-ray emission 2-3
times the normal average value\footnote{ The Crab steady state
(pulsar plus nebula) flux above 100 MeV detected by AGILE is
$F_{\gamma, steady} = (2.2 \pm 0.1) \times 10^{-6} \rm \, ph \,
cm^{-2} \, s^{-1}$.} until it reached on April 16, 2011 the
unprecedented high value of $F_{\gamma} = {\esn (19.6 \pm 3.7)}
\times 10^{-6} \rm \, ph \, cm^{-2} \, s^{-1}$ for a 24-hour {\esn
integration} \cite{atel-4} (hereafter $F_{\gamma}$ is the Crab
pulsar plus nebula flux above 100 MeV).
{\mtt This detection} of very rapid variations of the gamma-ray
emission
in the April 2011 flare confirms a
trend already noticed
 in the Fermi-LAT data {\mtt for} the September 2010 event by Balbo et al. 2011.
{\esn A re-analysis of the September 2010 event AGILE data, which
will be presented elsewhere, shows that the gamma-ray variability
on a timescale below 1 day is confirmed also by AGILE for that
event.}


The goal of our paper is threefold: (1) investigate the short
timescale structure of the gamma-ray emission of the April, 2011
event; (2) {\mtt present} the gamma-ray spectrum at the flare
peak;
(3) briefly discuss the theoretical implications of the very rapid
variability {\mtt and overall emission}.

\section{The April 2011 gamma-ray super-flare}

\begin{figure}
\begin{center}
\vspace*{-0.3cm} \hspace*{-0.7cm}
\includegraphics[width=10.5cm]{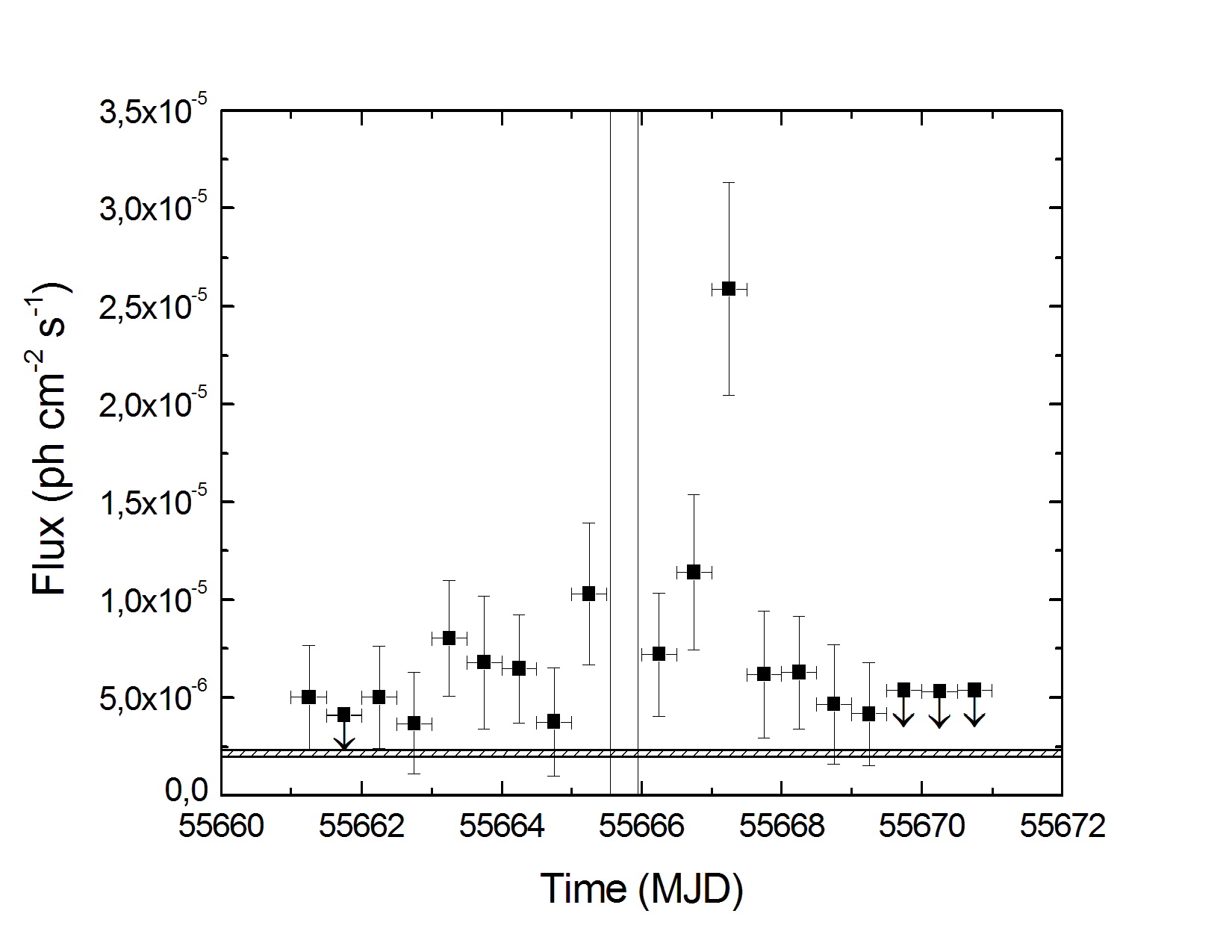}   
\caption{Crab (pulsar plus Nebula) gamma-ray 12-hr binned
lightcurve above 100 MeV detected by AGILE-GRID during the period
{\mtt 10-19} April, 2011. {\esn The gray horizontal band indicate
the Crab pulsar plus nebula average flux in the AGILE bandpass,
and the gray vertical lines mark the time period during which a
loss of AGILE-GRID telemetry data occurred because of ground
station activity. Peak emission occurred on April 16, 2011. } }
\label{fig-1}
 \end{center}
 \end{figure}

The AGILE satellite \cite{tavani2} has been monitoring in spinning
mode\footnote{The AGILE spinning mode allows a daily exposure of
about 70 \% of the sky depending on solar panel constraints, for
an instrument boresight rotation period of about 7 minutes.} the
Crab Nebula region with optimal exposure during the period
February-April, 2011. Starting on April 10-11, 2011 a noticeable
rising gamma-ray flux from a source positionally consistent with
the Crab Nebula\footnote{The possibility of a chance positional
coincidence with a background transient gamma-ray source has been
considered in T11 and A11. In light of the results presented also
in this paper, we consider this probability as negligible.} {\es
was recorded} by the Fermi-LAT instrument \cite{atel-1}. {\esn
Enhanced} emission with respect to the average flux of the steady
pulsar plus nebula emission  was detected at even larger flux
values by AGILE above 100 MeV during the following days, reaching
a value $F_{\gamma} = (6.5 \pm 1.5) \times 10^{-6} \rm\, ph \,
cm^{-2} \, s^{-1}$ on April 11-13, 2011 \cite{atel-2}. Remarkably,
the gamma-ray flux increased even more in the following days,
reaching the 1-day averaged flux of $F_{\gamma} = (12.1 \pm 0.6)
\times 10^{-6} \rm \, ph \, cm^{-2} \, s^{-1}$ on April 14, 2011
{\esn \cite{atel-3}}, and the even larger and currently
record-breaking 1-day averaged flux of $F_{\gamma} = (19.6 \pm
3.7) \times 10^{-6} \rm \, ph \, cm^{-2} \, s^{-1}$ on April
15-16, 2011 \cite{atel-4}. Fig.~\ref{fig-1} shows the {\mtt AGILE}
12 hr-binned gamma-ray lightcurve\footnote{ We note that a loss of
AGILE-GRID telemetry data occurred on April 14, 2011 caused by
pre-determined satellite tracking activity of the AGILE ground
station in Malindi (Kenya). This interval partially overlaps with
the first strong gamma-ray flare detected on April 14, 2011.}
(Crab pulsar plus nebula) during the period {\mtt 10-19} April,
2011
as obtained by the standard AGILE maximum likelihood procedure.

{\esn The 12-hr integration between MJD=55667.0 and MJD=55667.5
{\mtt (April 16, 2011)} yields for the peak flux the value
$F_{\gamma, P} = {\esn (26 \pm 5)} \times 10^{-6} \rm \, ph \,
cm^{-2} \, s^{-1}$, with a pre-trial statistical significance
$\sigma_{P}=10$. Considering the {\mtt accumulated} AGILE exposure
on the Crab Nebula in spinning mode (equivalent to 850 maps of
12-hr integrations), the post-trial significance for the super
flare turns out to be $\sigma_{P, post-t}=9.2$.} The statistical
significance of this 12-hr excess emission compared to the steady
Crab Nebula flux is {\esn 7.5} $\sigma$. {\es This significance is
calculated by the likelihood method taking into account the
standard average flux and position of the Crab pulsar and Nebula,
together with the catalogued gamma-ray sources in the field. {\mtt
After taking into account the accumulated exposure}, we find that
the post-trial significance of the excess emission is 6.4
$\sigma$. }
For a few days in April, 2011, the Crab
Nebula became the brightest gamma-ray source in the sky, rivaling
in intensity not only the Vela pulsar, but also the brightest
blazars ever detected in gamma-rays.

{\mtt In order to determine the Crab super-flare peak time and
duration,
we proceeded as follows. Starting from the 12-hr binned lightcurve
showed in Fig. 1, we obtained different lightcurves by
progressively shifting their bin starting times by 1 hour.
These 12 lightcurves\footnote{These datapoints were obtained by
the AGILE maximum likelihood procedure.} have been compared with
those
calculated from a flux model obtained by integrating for each bin
the model function  $$ f=\left \{\begin{array}{ll} B + A
e^{-(|t-t_{P}|/\sigma_{r})^{\nu}} \hbox{  for  } t\leq t_{P}
\\B + A e^{-(|t-t_{P}|/\sigma_{d})^{\nu}} \hbox{  for  } t> t_{P}
\end{array}
 \right .
$$ where $B=1 \times 10^{-5} \rm \, ph \,
cm^{-2} \, s^{-1} \, day^{-1}$ is the flux baseline (i.e., the
average of the 12 light curves baselines excluding the peak
emission), $t_p$ is the peak time, $\sigma_{r}$ and $\sigma_{d}$
the rise and decay time constants, respectively, and
 $\nu$ a measure of the pulse sharpness ($\nu =1,2$ for
two-sided exponential and gaussian fits, respectively; see, e.g.,
Norris et al. 1996).
In general, lower values of
$\nu$ imply a more peaked pulse. We also notice that the rise and
decay times, half to maximum amplitude, are obtained as
$\tau_{r,d} = [\ln(2)]^{1/\nu} \, \sigma_{r,d} $.
%

In the following, we conservatively assume a {\mttt Gaussian model
for the flaring} with $\nu = 2$, and $ \sigma_{r} = \sigma_{d}$.
From the values of $\sigma_{r,d}$ obtained by the best fit of each
lightcurve (for reduced $\chi^2$ values ranging from 1.2 and 1.4),
we derived that {\mttt the super-flare duration $ \sigma_{r} +
\sigma_{d}$ }
is constrained in the
{\mttt 68\% }   confidence level range between 14 hr and 26 hr.
%
%
Taking into account the relation $\tau_{r} = [\ln(2)]^{1/2} \,
\sigma_{r} \simeq 0.8 \, \sigma_{r}$,
we obtain the risetime range $ 6 \, {\rm hr} < \tau_r < 10 \, \rm
hr$.
This result is independent of the lightcurve bin zero-phase
($t_{start}$) choice. The value of the peak time $t_{P}$ is
determined as MJD=$55667.3\pm0.3$. }

\begin{figure}
\begin{center}
\vspace*{-0.3cm} \hspace*{-0.7cm}
 \includegraphics[width=10.5cm]{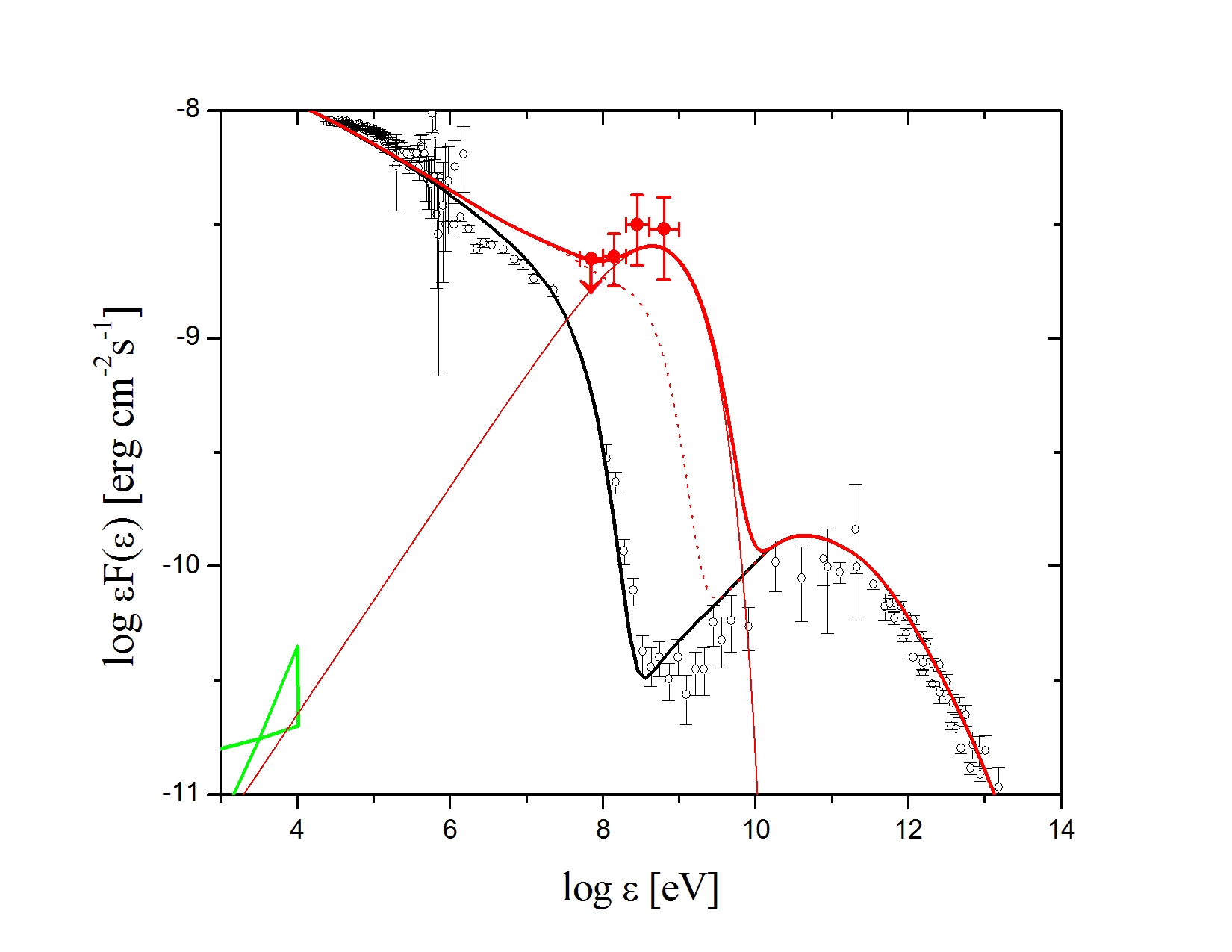}
\caption{The AGILE-GRID gamma-ray pulsar-subtracted spectrum of the Crab Nebula
super-flare on April 15-16, 2011.
The AGILE flaring spectral data, marked in red, obtained for a
1-day integration {\mtt (MJD $=$ 245566.4 -- 245567.4)}; data
points marked in black show the average nebular spectrum (Meyer et
al. 2010). {\mt Pulsar gamma-ray spectral data have been
subtracted based on the AGILE results presented in Pellizzoni et
al. (2009).} The red curve is the result of the theoretical
modelling of the super-flare as discussed in the text. {\mtt  The
spectral region marked in green shows the X-ray spectrum  of
``source
 A" which is the most dominant source in the \textit{Chandra} image of the ``anvil" region in the inner Crab Nebula as reported in T11}.
{\mt This flux level is indicative of an X-ray upper bound
expected from the flare. }} \label{fig-2}
 \end{center}
 \end{figure}

%

A complete spectral evolution of the April 2011 event will be
reported elsewhere. Here we focus on the 1-day integrated April
15-16, 2011 super-flare spectrum that we show in Fig.~\ref{fig-2}.
Very intense and relatively hard emission is detected in the
energy range 100 MeV - 1 GeV. The AGILE optimal spectral
sensitivity in the 50 MeV - a few GeV energy range is important in
constraining the spectrum at relatively low gamma-ray energies.
Indeed, the 50-100 MeV flux is well constrained by our 95\%
confidence level upper limit.  The super-flare emission shows a
very prominent peak of the $\nu F_{\nu}$ spectrum at photon
energies $E_{\gamma,P} \simeq 500 $~MeV. No significant emission
is detected above 1 GeV.

Fig.~\ref{fig-2} shows also the results of our theoretical
modelling of the emission (red curve) that we discuss below.
Remarkably, the peak power emitted in the gamma-ray energy range
between 100 MeV and a few hundreds of MeV equals the average power
emitted in the hard X-ray/MeV energy range by the Crab Nebula.

\section{Theoretical constraints}

In modelling {\es the} April 2011 Crab Nebula gamma-ray event we assume that
fast and very efficient acceleration is occurring at a site in the
inner nebula, following the discussion of T11 and Vittorini et
al., 2011 (hereafter V11). A fraction of the total
electron-positron high-energy component in the Nebula is
impulsively accelerated at a site of size $L$. For simplicity, we
ignore here {\mt substantial} enhancements due to Doppler
boosting, { \mt and take the  Doppler factor ${\delta} = \Gamma^{-1}(1 -
\beta \cos \theta)^{-1} $ to be of order of a few.}


{\ess The physical quantities are constrained  within a global comparison
of a multi-parameter model matching spectral and timing data\footnote {{\ess Having
determined from the overall spectral shape the values of $\gamma_b$ and index $s$
(see their definitions in the text), we have five remaining parameters:
$\gamma_{max}$, the local magnetic field $B$, the electron density $N_e$, the
dimension of the emitting region $L$, and the Doppler factor $\delta$ (the parameter
$K$ is derived from $N_e$ and $L$).
These parameters are obtained from the following quantities (in the observer frame):
the position of the peak emission, $ E_p \propto\delta \, \gamma_{max}^{2} \, B$,
the peak emission $\nu F \propto \delta^4 \,  N_e \, L^3 \, B^2 \, \gamma_{max}^2$,
the rise time $\tau_r=L/(c\,\delta$), and
the cooling time $\tau_c \propto 1/(B_{loc}^2\, \gamma_{max} \, \delta)$.}}.
We considered several models with the assumption of $\delta$ in the range $1 - 4$ as deduced from
observations of the South East jet and wisp regions (e.g. Hester 2008).
We present here the cases with $\delta=1$ and $\delta=4$ as examples of a class of models
applied to the} super-flare
{\mt spectrum shown in Fig. 3}.

The acceleration process produces, within a timescale shorter than any other
relevant timescale,
 a {\mt differential } particle energy distribution (that we model for illustration purposes
 in its simplest form as a single power-law
 distribution
$ dn/d\gamma=K \, \gamma_b^{-1}/(\gamma/\gamma_b)^{s} $
\noindent where $n$ is the {\mt local} particle number density,
$\gamma$ is the particle Lorentz factor  ranging from
$\gamma_{min}=10^5$ to $\gamma_{max}=7\times 10^{9}$,  $s=2$  is
the power-law index, $\gamma_b = 5 \times 10^8$, and $K$ is the
normalization factor $K = {\esn 4}\times 10^{-7}$~cm$^{-3}$.
{\ess For $\delta=1$, the emitting region has size $L = 10^{15}\, $cm,
and an enhanced local magnetic field
$B_{loc} = 2 \times 10^{-3} \,$G that we keep constant in our
calculations.} The total particle number required to explain the
flaring episode turns out to be $ N_{e-/e+} = \int dV \,
(dn/d\gamma) \, d\gamma \simeq 7\times 10^{42}$, where $V$ is an
assumed spherical volume of radius $L$. {\mt{\ess For $\delta = 4$},
some of the physical parameters are slightly
different, e.g., $\gamma_{max}=5\times 10^{9}$, $B_{loc} = 1.3
\times 10^{-3} \,$G, $K = {\esn 3} \times 10^{-10}$~cm$^{-3}$, $L =
{\esn 4} \times 10^{15}\, $cm, and $ N_{e-/e+} = 3\times 10^{41}
$. Obviously, the physical parameters can differ from
these (and are even more extreme) for time variations faster than
the 1-day spectral average of {\mttt Fig. 2.} A  discussion of the
complete lightcurve and physical implications will appear
elsewhere.}
The complex  gamma-ray lightcurve shows that the acceleration
process occurs on a $\sim$ week timescale with a succession of
short timescale flares, each of which has physical parameters
which differ from those of the peak emission by a factor of a few.

%

 We find that the synchrotron
peak photon energy during the flare maximum  is $ E _{peak } =
\frac{3}{2} \, \hbar \, \frac{e \, B_{loc}}{m_e \, c} \,
\gamma_{max}^2 \simeq 500 \; \rm MeV   $, a value that challenges
models of diffusive particle acceleration limited by synchrotron
cooling, a fact already noticed in T11, A11, and V11. The April
2011 event confirms even more the extremely short timescale of
acceleration occurring in the inner Crab Nebula, and
the existence of a strongly enhanced local magnetic field.




\section{Discussion and Conclusions}

The April 2011 Crab Nebula super-flare dramatically shows the
efficiency of the particle acceleration mechanism operating in the
inner nebula. {\mttt The detected gamma-ray luminosity at the peak
of the April 16, 2011 super-flare corresponds to 0.3\% of the Crab
pulsar spindown luminosity.} Despite recent high-resolution
observations of the inner nebula (see especially the very
interesting sequence of Chandra pointings reported by Tennant et
al. 2011), there is currently no identification of the
acceleration site. The anvil region has been suggested as a
candidate for the September 2010 event (T11), and this site may
well be active also in the case of the April 2011 event. More
observations and extended monitoring of the Crab Nebula is
necessary {\mtt also in light of the "secular" X-ray variations as
determined over a timescale of years \cite{wilson-hodge}}.

The challenge to particle acceleration models applies to the main
theoretical frameworks that have been so far proposed for the Crab
Nebula: the magnetohydrodynamical approach of Kennel \& Coroniti
(1984), diffusive shock acceleration mechanisms (e.g., Drury 1983,
Blandford \& Eichler 1987), shock drift acceleration (e.g., Kirk
2000, Reville \& Kirk 2010), magnetohydrodynamical instabilities
(e.g., Komissarov \& Lyubarsky 2004, Del Zanna et al. 2004, Camus
et al. 2009, Komissarov \& Lyutikov 2010), ion-driven acceleration
(e.g., Arons 2008). The data presented here contribute in a
substantial way to a step further in deepening our understanding
of acceleration processes, and {\es may force the previously proposed
models to be substantially revised}. Of particular relevance is
the possible role of impulsive particle acceleration in magnetic
field reconnection, and/or runaway particle acceleration by
transient electric fields violating the condition $E/B < 1$ (that
is  typically assumed in standard models, e.g., deJager et al.
1996). {\ess The super-flare spectrum of Fig. 2, showing significant emission
above $200$ MeV, implies $E/B \ge 2$ \cite{tavani5}.}
The applicability of these concepts to the Crab Nebula
flaring activity remains to be tested by future investigations.


Research partially supported by the ASI grant no. I/042/10/0.



\end{document}